**De l'imagerie médicale à la réalité virtuelle pour l'archéologie**

Théophane Nicolas, Ronan Gaugne, Bruno Arnaldi et Valérie Gouranton

**Les limites de la numérisation pour l'archéologie**

La culture matérielle et le lien à son contexte historique ainsi que son environnement sont au cœur du métier de l'archéologue. Mobiliers archéologiques et sédiments sont des témoignages fragiles qu'il convient d'analyser, d'interpréter, de préserver et de valoriser. Cette relation complexe est en pleine mutation à l'ère du numérique, au moment où l'archéologue a accès à des représentations virtuelles de l'objet, voire à une copie physique obtenue par impression 3D. Ces dernières années ont vu l'essor de méthodes de numérisation (photogrammétrie…) qui apportent une réponse pertinente en fournissant des solutions non destructives de conservation, d'analyse, et de transmission du patrimoine archéologique mobilier et immobilier. Cependant, ces techniques sont limitées à la surface visible des objets, monuments ou sites. Or, les informations structurelles (non limitées à leurs surfaces) et contextuelles (liées à leur relation à l'environnement) du matériel archéologique représentent des aspects primordiaux pour l'archéologue, auxquels le numérique peut également apporter des éléments de réponse. Pour pallier ces limitations, les technologies d'imagerie médicale sont également de plus en plus utilisées dans le domaine de l'archéologie, car elles permettent un accès non destructif à la structure interne d'artefacts* souvent fragiles. Toutefois, et la plupart du temps, cette utilisation reste limitée à une simple visualisation et porte généralement sur des pièces exceptionnelles (momies, artefacts de grande valeur). Les données obtenues par les technologies d'imagerie telles que la tomodensitométrie* (qui fournit des images en coupes d'un objet) sont constituées d'éléments d'information retranscrits de manière visuelle, mais dont la richesse intrinsèque peut être exploitée de manière plus large grâce à des technologies issues des dernières recherches en 3D, telles que la réalité virtuelle*, la réalité augmentée*, les interactions* et la fabrication additive*. Le développement de ces technologies permet aujourd'hui d'envisager une démocratisation des procédés, tout particulièrement dans le domaine de l'archéologie préventive où les perspectives ne se limitent pas uniquement à la valorisation.

Le projet IRMA (Imageries et inteRactions Multi-modales pour l'Archéologie) propose des méthodologies innovantes pour la recherche dans le domaine du patrimoine historique et archéologique basées sur une combinaison de ces technologies d'imageries médicales et de modalités de restitutions 3D interactives, en s'appuyant sur des résultats de recherche issus d'une collaboration entre l'IRISA (Institut de recherche en informatique et systèmes aléatoires), le CReAAH (Centre de Recherche en Archéologie, Archéosciences, Histoire), l'Inrap (Institut de recherche en Archéologie

Préventive) et l'entreprise Image ET. Ces nouvelles méthodologies sont destinées aux professionnels du patrimoine culturel, tels que les musées, les conservateurs, les restaurateurs et les archéologues.

**Au-delà de la numérisation, l'interaction en réalité virtuelle**

En combinant différents types de technologies d'acquisition numérique issues de l'imagerie médicale, il est possible d'identifier et d'analyser scientifiquement, par des méthodes non destructives et efficientes, des objets archéologiques non visibles (comme des ossements brûlés dans une urne en céramique), d'évaluer la fragilité, l'état de conservation et la possibilité de restauration d'un artefact archéologique corrodé, de visualiser, analyser et manipuler physiquement des objets inaccessibles et/ou fragiles (tomodensitométrie, impression 3D), et ainsi de révéler notre patrimoine culturel caché et de l'observer dans son contexte (réalité virtuelle, augmentée ou mixte, 3D).

La tomodensitométrie combinée aux différentes modalités 3D permet de visualiser l'intérieur des artefacts et objets et d'accéder à différents éléments intriqués. Le rendu volumique obtenu par reconstruction 3D de cette dernière permet d'associer des colorations et des niveaux de transparence aux différentes valeurs de densité, donnant ainsi un premier niveau de visualisation spatialisé sur écran. En y associant la fonctionnalité de segmentation*, il devient possible de se concentrer sur certaines parties spécifiques de l'objet à étudier et d'en extraire virtuellement un élément. La génération d'un maillage 3D à partir des informations de densité autorise la production de copies physiques, soit d'éléments particuliers, soit de l'objet étudié dans son intégralité.

Les données 3D ainsi générées peuvent ensuite être intégrées dans des applications interactives, en réalité virtuelle ou augmentée. Ces applications permettent d'associer un contexte aux objets numérisés, qui peut être technique, comme un contexte de fouille virtuelle, ou fonctionnel, comme un contexte d'utilisation ou de fabrication de l'objet. Dans le cadre du projet IRMA, la méthodologie mise en œuvre a été utilisée dans différents cas d'étude qui nous ont permis d'en illustrer et d'en démontrer l'intérêt selon deux principaux usages : la visualisation et la manipulation.

**Mieux visualiser et manipuler**

L'étude de l'urne (datant du 1$^{er}$ ou 2eme âge du fer) de Guipry (en Ille-et-Vilaine) est caractéristique de ce double usage recherché par les archéologues. Grâce à la tomodensitométrie (figure 1), son contenu a pu être visualisé très rapidement et ses différents composants ont été identifiés : des fragments d'os et deux objets en métal, une lame de couteau et une fibule* (sorte d'épingle ou de fermoir en métal). La fibule a pu être retirée virtuellement de l'amas, par segmentation, tout comme la gangue de corrosion qui la recouvrait. Cette opération a permis de générer un modèle 3D de cette fibule et d'obtenir par la suite une impression 3D de l'objet bien avant son extraction physique. Une impression complète de l'urne en transparence permet de conserver une représentation tangible de l'organisation spatiale interne de l'urne après sa destruction (figure 2).

L'impression 3D permet de manipuler l'objet sans risque de dommage pour l'original. Nous avons illustré l'intérêt de cette technique par une copie démontable d'un poids gaulois qui a permis de mieux comprendre sa composition et les techniques de fabrication, par des copies à différentes échelles d'objets trop petits pour être manipulés, ou encore par des copies d'objets dont il ne restait que le négatif (figure 3). Combinée à la réalité virtuelle, la copie d'un objet peut servir d'interface aidant à mieux comprendre le contexte d'utilisation de l'original. Nous avons ainsi proposé une reconstitution d'une balance romaine avec manipulation interactive du poids, présentée dans le cadre d'une action de médiation scientifique (figure 4).

**Des changements de paradigme pour l'archéologue**

Ces outils offrent à l'archéologue de nombreux avantages. Ils lui permettent, par une méthode non destructive, d'exploiter scientifiquement des objets fragiles et/ou non visibles avant une intervention préventive préalable (conglomérat d'objets, objet oxydé) ou une fouille (amas cinéraire dans une urne), en visualisant un modèle numérique ou bien en travaillant sur le fac-similé réalisé à partir d'une impression 3D plutôt qu'en manipulant l'objet original. Ils rendent possible la visualisation, l'analyse et la manipulation d'objets archéologiques sous forme de modèles numériques en coupe ou en trois dimensions, avec possibilité de prise de mesures. La mise en œuvre d'un modèle 3D permet un archivage numérique pérenne de l'objet, et ce quelles que soient les interventions ultérieures (fouille dans le cas d'un contenant ou d'un conglomérat), pouvant être visualisé et analysé à distance par plusieurs opérateurs. La tomodensitométrie aide à déterminer la densité du ou des matériaux, à évaluer la fragilité ou l'état de conservation des objets archéologiques et à orienter les choix de traitements.

Malgré la démocratisation de ces outils et leur développement rapide, il subsiste toutefois des problèmes d'accessibilité à l'appareillage et des contraintes techniques et technologiques. Concernant l'acquisition *via* un scanographe, la matière ou la géométrie de l'objet peuvent engendrer des artefacts qui nuisent, voire empêchent une lecture correcte. Notre expérience montre également toute l'importance de la capacité de réglage des constantes du scanographe, que seul du matériel industriel permet. Le second paramètre clé est la capacité de l'opérateur à faire de la reconstruction 3D, afin d'obtenir des images optimales. La tomodensitométrie offre une précision limitée aux caractéristiques physiques du capteur ; pour pallier cet inconvénient, il est possible de faire appel à la micro-tomodensitométrie qui augmente la précision d'analyse. Toutefois, si la tomodensitométrie permet d'acquérir des images sur des objets de grande taille, ce n'est pas le cas de la micro-tomodensitométrie, qui est réservée à des éléments de petites dimensions. Concernant l'impression 3D, les limitations sont plutôt d'ordre technologique : précision de l'impression elle-même, taille de l'objet fini, mais également contraintes sur le modèle 3D qui sert à faire l'impression et qui nécessite de fortes compétences techniques. L'usage de la réalité virtuelle est également limité par ces trois paramètres de l'accès à la technologie, des compétences requises et des limitations de l'appareillage. Se pose enfin la question de la position spécifique de l'archéologue qui emploie ces nouveaux outils, et de la manière dont ils modifient sa perception du mobilier et de son contexte.

**Conclusion**

Ces technologies ouvrent de nouvelles perspectives de recherches et de méthodologie. Elles se révèlent être un support et un soutien pertinent aux opérations de fouille ou de restauration, en remettant dans leur contexte des artefacts, et en y associant des interactions naturelles grâce à la réalité virtuelle ou augmentée. Mais, issues d'un milieu médical ou industriel, elles restent toutefois à adapter à des usages spécifiquement archéologiques. Quoi qu'il en soit, ces outils 3D sont porteurs d'un changement de paradigme pour l'archéologue et auront sans aucun doute un impact à l'avenir sur sa manière de percevoir le mobilier et son environnement.

**Figures**

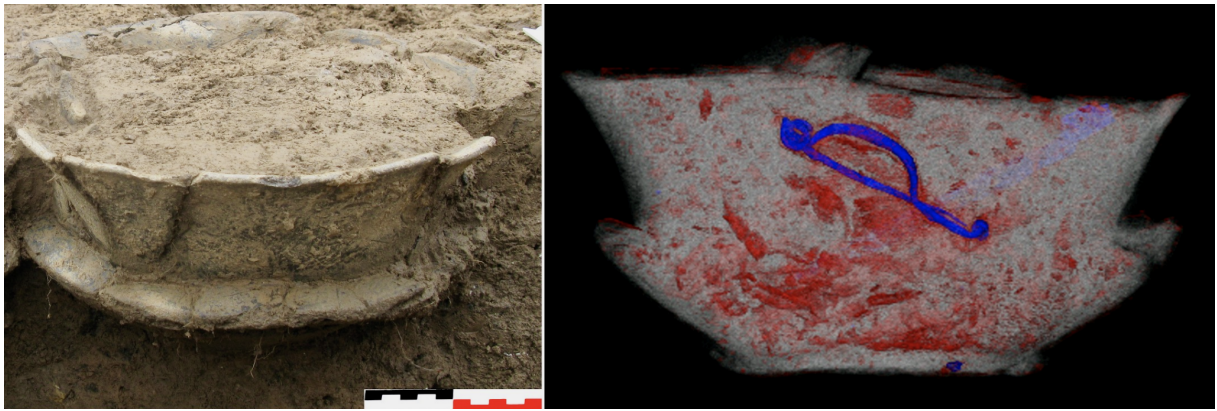

**Figure 1.** Gauche : Urne funéraire de Guipry (Ille-et-Vilaine). Droite : Tomodensitométrie de l'urne (les zones rouges correspondent aux ossements, en gris le sédiment et en bleu au objets métalliques).

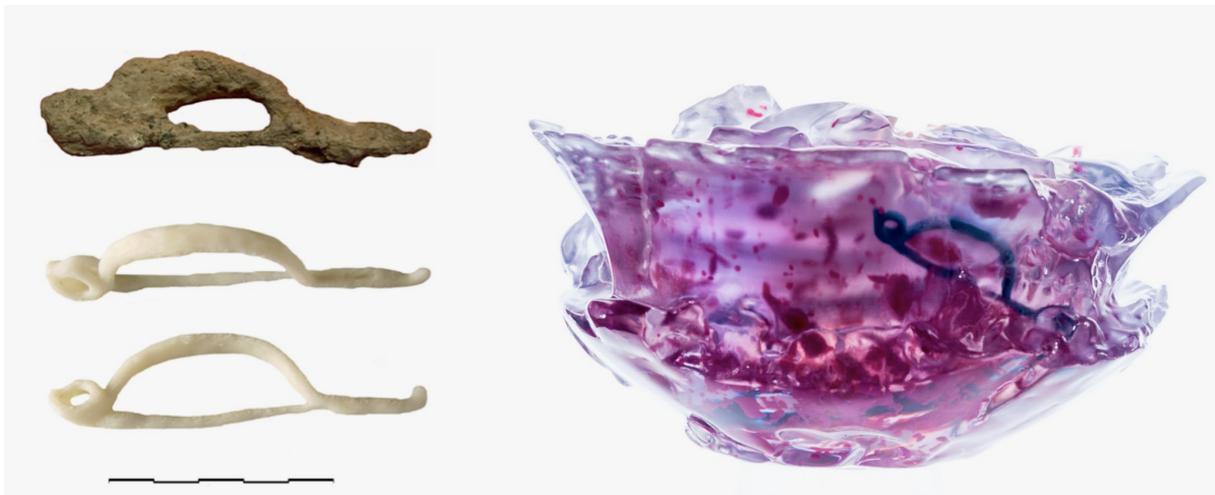

**Figure 2.** Gauche : La fibule extraite de l'urne et sa copie en impression 3D. Droite : Impression complète de l'urne en transparence.

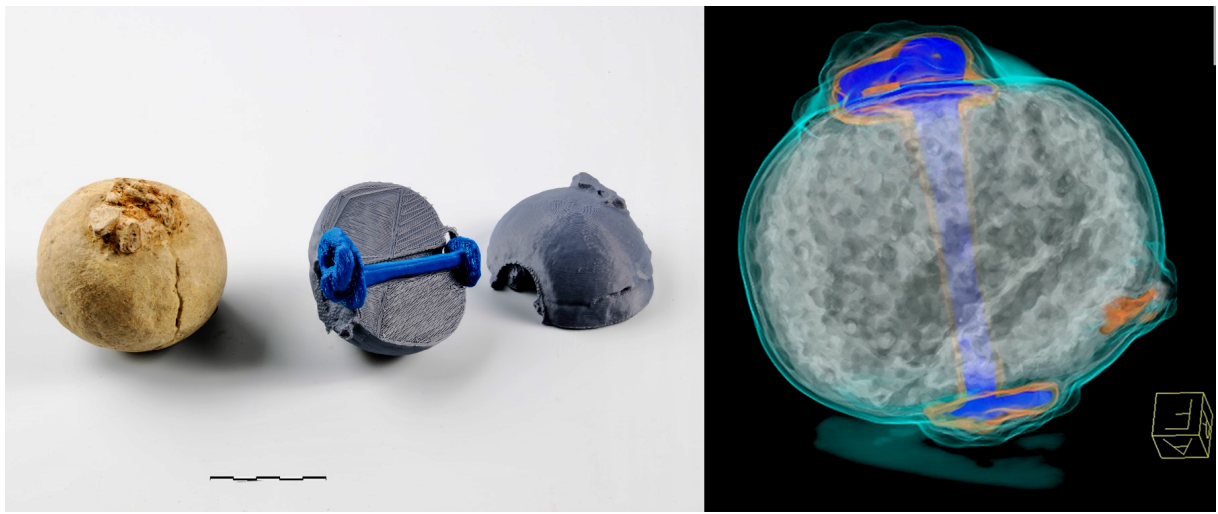

**Figure 3.** Gauche : Poids gaulois du site d'Ossé (Ille-et-Vilaine) et sa copie démontable en impression 3D. Droite : Tomodensitométrie du poids (les zones grises correspondent à la pierre granitique, les zones bleues au métal, les zones oranges, à la gangue de corrosion, les zones turquoises à du sédiment.).

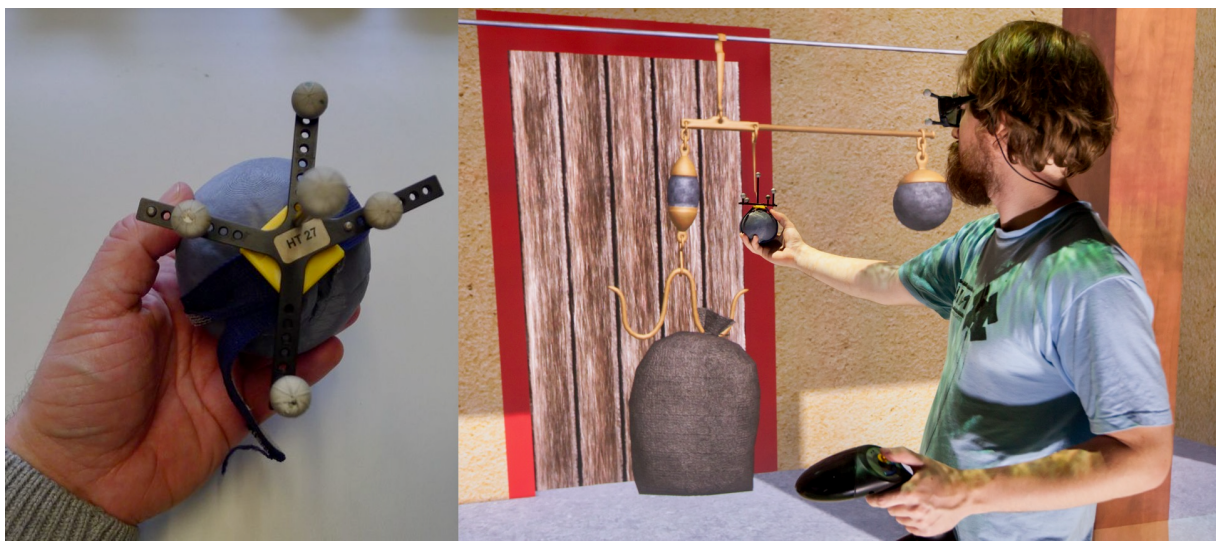

**Figure 4.** Gauche : Impression 3D du poids avec marqueur de capture de position. Droite : Interaction en réalité virtuelle avec la reconstitution de la balance.

**Glossaire**

**Artefact.** Vestige archéologique façonné par l'être humain et découvert à l'occasion de fouilles archéologiques.

**Fabrication additive.** La fabrication additive ou impression 3D est un procédé de création d'objets tridimensionnels tangibles à partir d'un fichier numérique.

**Fibule.** Une fibule est une « agrafe » généralement en métal, qui sert à fixer les extrémités d'un vêtement.

**Interaction.** L'interaction avec les objets du monde virtuel : l'utilisateur peut attraper les objets, les manipuler, les mettre en fonction ; les caractéristiques des objets sont modifiées et font partie de la modélisation du monde virtuel, dans le but de simuler un monde réel.

**Réalité virtuelle.** La réalité virtuelle (RV) est un domaine scientifique et technique exploitant l'informatique et les dispositifs d'interaction en vue de simuler, dans un environnement virtuel, le comportement d'entités 3D, qui sont en interaction en temps réel entre elles et avec un ou plusieurs utilisateurs en immersion pseudo-naturelle par l'intermédiaire de canaux sensorimoteurs.

**Réalité augmentée.** La réalité augmentée (RA) est un domaine des applications informatiques qui combine le réel et le virtuel, en temps réel et donne l'apparence que les objets virtuels et réels cohabitent dans le même monde tridimensionnel. Elle augmente la visualisation de la réalité avec des données numériques et permet à l'utilisateur d'interagir avec l'environnement

**Segmentation.** Suite à une tomodensitométrie, il est possible de mener une opération qui consiste à extraire une ou des parties d'intérêt d'un objet. Cette partition virtuelle consiste à sélectionner manuellement ou automatiquement une ou des régions d'intérêts afin de construire des modèles 3D séparés. Cette technique permet de dégager virtuellement un objet.

**Tomodensitométrie.** La tomodensitométrie est le nom scientifique de la scanographie par rayons X. Son but est de permettre l'acquisition axiale d'un objet et de le soustraire à la problématique de superposition des éléments internes de l'objet scanné qui peuvent provoquer des occlusions et donc des pertes d'information. Cette acquisition fournit des images en coupe d'un objet à trois dimensions dont le cumul est utilisé par un logiciel dédié pour générer une reconstruction 3D.

**Affiliations**

**Théophane Nicolas.** Archéologue, Inrap, UMR Trajectoires, Cesson-Sévigné. Theophane.Nicolas@inrap.fr

**Ronan Gaugne.** Ingénieur de recherche en informatique, Univ. Rennes, IRISA, Inria, Rennes. Ronan.Gaugne@irisa.fr

**Bruno Arnaldi.** Enseignant-Chercheur en informatique, Univ. Rennes, INSA Rennes, IRISA, Inria, Rennes. Bruno.Arnaldi@irisa.fr

**Valérie Gouranton.** Enseignante-Chercheure en informatique, Univ. Rennes, INSA Rennes, IRISA, Inria, Rennes. Valerie.Gouranton@irisa.fr